\documentstyle[11pt,fleqn,epsf,a4wide]{article}

\newcommand{\Tr}{\makebox{ Tr }}

\newcommand{\GeV}{\makebox{ GeV}}

\newcommand{\beq}{\begin{equation}}

\newcommand{\enq}{\end{equation}}

\newcommand{\beqa}{\begin{eqnarray}}

\newcommand{\enqa}{\end{eqnarray}}

\newcommand{\nn}{\nonumber}

\newcommand{\lbq}[1]{\label{#1} \enq}

\newcommand{\lbqa}[1]{\label{#1} \enqa}

\newcommand{\befi}[1]{\begin{figure}[ht] \leavevmode \centering 
\epsffile{#1.eps}}

\newcommand{\lbfi}[1]{\label{#1} \end{figure}\parindent0em}

\newcommand{\eq}[1]{eq.(\ref{#1})}

\newcommand{\fig}[1]{fig.(\ref{#1})}

\newcommand{\ct}{\cite}

\newcommand{\pa}{\partial}

\newcommand{\cP}{\mbox{$\cal P$}}

\newcommand{\cS}{\mbox{$\cal S$}}

\newcommand{\bA}{\mbox{\bf A}}

\newcommand{\bF}{\mbox{\bf F}}

\newcommand{\bt}{\mbox{\bf t}}

\newcommand{\bW}{\mbox{\bf W}}

\newcommand{\bo}{\mbox{\bf 1}}

\newcommand{\al}{\alpha}

\newcommand{\de}{\delta}

\newcommand{\ep}{\epsilon}

\newcommand{\la}{\lambda}

\newcommand{\rh}{\rho}

\newcommand{\si}{\sigma}

\newcommand{\ch}{\chi}

\newcommand{\De}{\Delta}

\newcommand{\Ps}{\Psi}

\begin{document}
\setcounter{topnumber}{3} \setcounter{totalnumber}{3} \sloppy
\title{\begin{flushright}\normalsize HD-THEP-96-04\end{flushright}
  \LARGE \bf Nucleon Structure and High Energy Scattering} \author{
  Michael Rueter\thanks{supported by the Deutsche
    Forschungsgemeinschaft}, H.~G.~Dosch\\[.7cm] \it Institut f\"ur
  theoretische Physik\\ \it Universit\"at Heidelberg\\ \it
  Philosophenweg 16, D-69120 Heidelberg, FRG\\[.3cm] \it e-mail:
  M.Rueter@ThPhys.Uni-Heidelberg.DE } \date{} \maketitle
\thispagestyle{empty}
\begin{abstract}
  If the pomeron is generated by a two gluon exchange there is no a
  priori reason for a drastic suppression of three gluon exchange with
  negative parity and charge parity. This would lead to an
  unacceptably large difference between $p\,p$ and $p\, \bar{p}$
  scattering. It is shown that a natural suppression of the $C$=$P$=--1
  contribution to high energy scattering is given by a cluster
  structure of the nucleon.
\end{abstract}
\newpage \mathindent0em \setcounter{page}{1} {\Large \bf 1
  Introduction}\hspace{1cm}The possibility that the real part of the
scattering amplitude increases with energy as fast as the imaginary
part was first considered by Lukaszuk and Nicolescu \ct{1}. Such a
behavior which is well compatible with our present knowledge of
axiomatic field theory would mean that a trajectory of a pole which is
odd under C and P has an intercept near one, it has been called
odderon. One consequence of such an odderon would be that the ratio of
the real to imaginary part of the forward scattering amplitude is
different for particle particle and particle anti-particle scattering
even at asymptotic energies. For further reference we shall use the
conventional abbreviation $\De \rh$ for that difference: \beq \De
\rh(s) = \rh^{\bar p p}(s) - \rh^{p p}(s)=\frac{{\rm Re}\left[ T^{\bar
    p p}(s,0)\right]}{{\rm Im}\left[ T^{\bar p
    p}(s,0)\right]}-\frac{{\rm Re}\left[ T^{ p p}(s,0)\right]}{{\rm
    Im}\left[ T^{ p p}(s,0)\right]} \lbq{rhodef} Interest in the
odderon rose again when the UA4 collaboration \ct{2} reported a value
for $\rh^{\bar p p}$ at $\sqrt{s} = 546$ GeV which was much larger
than the one extrapolated by means of dispersion relations for
proton-proton scattering and a large value for$|\De \rh |$ seemed 
indicated.\\ The
new results of the UA4/2 collaboration \ct{3} obtained however a value
$\rh^{\bar p p}(\sqrt{s} = 541 \GeV)= 0.135 \pm 0.015$ which is very
well compatible with $\De \rh = 0$ at that energy (see e.g. \ct{4})
and at any rate leaves no room for a large value of that quantity. The
very successful description of high energy data by the
Donnachie-Landshoff pomeron \ct{LaPo} yields also $\De \rh \approx 0$.\\ 
The
new analysis of the $C$=$P$=--1 exchange had however shown that from
the point of view of QCD the odderon was by no means an odd concept.
In perturbative QCD it had been shown \ct{5} that the exchange of
three reggeized gluons leads in the leading log approximation of
perturbative QCD to an intercept of the $C$=$P$=--1 trajectory above
one, i.e.~its contribution is increasing with energy (though slower
than that of the perturbative pomeron). Similar results have also been
obtained in a non-perturbative approach using the N/D method \ct{6}.\\ 
As far as the contribution of three {\em non-perturbative} gluons is
concerned there is no reason for a strong suppression of the three
gluon versus the two gluon exchange. In an Abelian model for
non-perturbative gluon exchange \ct{7} Donnachie and Landshoff \ct{8}
have found that the lowest order effective odderon coupling, i.e.~the
coupling of three non-perturbative gluons, is suppressed by a factor
of two with respect to the effective pomeron coupling. Though it is
very gratifying that in a non-perturbative model the three gluon
coupling is smaller than the two gluon coupling (naive expectation
goes in the opposite direction), this coupling still leads to a value
of $|\De \rh |\approx .5$ which is far from consistent with the analysis
of the data. In the Abelian model of Landshoff and Nachtmann, where
quark additivity is a consequence of the model, the $\rh$-parameter for
hadron-(anti)hadron scattering is just the one for quark-(anti)quark
scattering.\\ In a series of papers (\ct{9} and the literature quoted
there) a non-Abelian model of high energy scattering was presented
which gives a good description of the data and relates parameters of
high energy scattering to those of hadron spectroscopy. One of the
most characteristic features of this model is that the same mechanism
which leads to confinement introduces a kind of string-string
interaction in high energy scattering and leads to a marked increase
of the total cross section as a function of the hadron size, even if
the latter is large as compared to the gluonic correlation length.
Quark additivity does not hold in that approach. The different total
cross sections for pion-nucleon, kaon-nucleon and
nucleon-(anti)nucleon scattering are correctly reproduced due to the
different (electromagnetic) radii of the hadrons. In this note we
evaluate the leading $C$=$P$=--1 contribution of that model. We show
that this contribution (and therefore also $\De \rh$) depends
crucially on the structure of the nucleon; we especially discuss the
dependence of $\De \rh$ on the radius of a di-quark if two quarks are
clustered.\\ Our paper is organized as follows: In section 2 we
shortly refer to the main ingredients of the model of the stochastic
vacuum (for a detailed description we refer to \ct{9}) and calculate
the $C$=$P$=--1 contribution to the scattering amplitude. In section 3
we give the numerical results for $\De \rh$ for different nucleon
configurations and discuss the implications in section 4.

{\Large \bf 2 The color singlet $\bf C$=$\bf P$=--1 scattering
  amplitude}\\ The main ingredients for the above mentioned treatment
of non-perturbative high energy scattering are separation of the large
energy from the small momentum transfer scale by the eikonal
approximation for a fixed gluon vacuum field and subsequent averaging
over these fields \ct{10}; the averaging is done with the model of the
stochastic vacuum (MSV) \ct{11}\ct{12}. In our non-Abelian treatment
it is crucial to respect gauge invariance and hence the fundamental
processes are not quark-quark scattering, but rather
,,scattering'' of Wegner-Wilson loops (see \fig{2loops}). For details
we refer to \ct{9} and only shortly indicate in words and figures the
principal steps of the procedure.

As mentioned the principal ingredient is the scattering amplitude of
two Wegner-Wilson loops with light-like sides. The line integrals
$\exp \{-ig \int \bA dz\}$ occuring along the light-like sides of the
loops are just the eikonal phases of the constituents. In order to
evaluate the amplitude, i.e.~to perform the functional integration
over the gluon fields, first the line integrals over the potentials
are transformed into surface integrals over the field strengths by 
means of the non-Abelian Stokes theorem \ct{NonAb}. In
doing so one has to introduce a reference point $C$ which is common to
both the surfaces bordered by the loops (see \fig{2pyramiden}). The
expectation value of the two loops is then evaluated in the model of
the stochastic vacuum which assumes that the non-perturbative gluonic
contribution can be approximated by a Gaussian stochastic process in
the field strengths $\bF_{\mu\nu}$. This Gaussian process is
characterized by the two-point function of two parallel transported
field strengths expressed in the adjoint basis of the $SU(3)$ \beq
<F^A_{\mu\nu}(x,w)F^B_{\rh\si}(y,w)>.  \lbq{Korrelator} Here $w$ is
the coordinate of the reference point $C$ mentioned above (see
\fig{2pyramiden}). All higher correlators can be reduced to products
of this two-point function for which the MSV makes a Lorentz- and
gauge-invariant ansatz.

If we consider meson-meson scattering, the loops shown in \fig{2loops}
must be averaged with a transversal meson wave function.\\ In order to
describe baryon-baryon scattering, one has to start from three loops
(without traces) with one common side as shown in \fig{3loops}. From
these loops a baryon is constructed by averaging with a transversal
wave function. We consider two classes of configurations:\\ In the
first case the distances from the common line of all three loops are
equal and we vary the angle $\al$ between two loops (see
\fig{3loops}). If this angle tends to zero the two quarks together
form a point-like di-quark (i.e.~an object transforming as the $\bar
3$ representation of $SU(3)$).\\ The other case we consider is a
linear structure of the nucleon.

The light-like components of the surface integrations of the
Wegner-Wilson loops can be performed analytically and we end up in the
transversal plane of the scattering process. In the following the
transversal components of a vector are denoted by
$\vec{x}=(0,x^1,x^2,0)$.\\ It is convenient to introduce a reduced
scattering amplitude for baryon-baryon scattering \ct{9} depending on
the impact parameter $\vec{b}$ and the extension parameters
$\vec{R}_i$ of the two baryons: \beqa
\tilde{J}(\vec{b},\vec{R}_1,\vec{R}_2)&=&-<B_1\cdot B_2>\nn\\ {\rm
  with }\;B_i&=&\frac{1}{6}\ep_{abc}\ep_{a'b'c'}\left\{
W_{a'a}[\cS_{i1}]W_{b'b}[\cS_{i2}]W_{c'c}[\cS_{i3}]-\de_{a'a}
\de_{b'b}\de_{c'c}\right\}.
\lbqa{Jtilde} The unitary $3\times 3$ matrices $\bW [\cS_{ij}]$ are
the Wegner-Wilson loops \beq W_{a'a}[\cS_{ij}]= \left[\cP\,
e^{-ig\oint_{\pa S_{ij}}\bA_\mu (z)\,{\rm d}z^\mu}\right]_{a'a}
\lbq{WWloopa} and the integration paths $\pa \cS_{ij}$ are illustrated
in \fig{3loops}.\\ Before we proceed further we want to show that from
this \eq{Jtilde} one can see easily that in the limit of the angle
$\al$ (see \fig{3loops}) going to zero, the baryon can effectively
treated like a meson; the resulting di-quark playing the role of the 
anti-quark.
If $\al \rightarrow 0$ the loop $\pa \cS_{i2}$ goes over into $\pa
\cS_{i1}$ and we have \beq B_i(\al =
0)=\frac{1}{6}\ep_{abc}\ep_{a'b'c'}\left\{
W_{a'a}[\cS_{i1}]W_{b'b}[\cS_{i1}]W_{c'c}[\cS_{i3}]-\de_{a'a}\de_{b'b}
\de_{c'c}\right\}.
\lbq{3a} We use the following identity for $SU(3)$ matrices
\[
\ep_{a'b'c'}W_{a'a}W_{b'b}W_{c'c}=\ep_{abc}
\]
from which we obtain \beq
\ep_{a'b'c'}W_{a'a}[\cS_{i1}]W_{b'b}[\cS_{i1}]=\ep_{abh}W^{-1}_{hc'}
[\cS_{i1}]=
\ep_{abh}W_{hc'}[\cS_{i1}^{-1}] \lbq{3b} where $\pa \cS_{i1}^{-1}$ is
the Wegner-Wilson loop oriented in opposite direction. Inserting
\eq{3b} in \eq{3a} we obtain \beqa B_i(\al =
0)&=&\frac{1}{3}\de_{ab}\left\{
W_{ac}[\cS_{i1}^{-1}]W_{cb}[\cS_{i3}]-\de_{ab}\right\}=\frac{1}{3}
\de_{ab}\left\{
W_{ab}[\hat{\cS}_{i12}]-\de_{ab}\right\}\nn\\ &=&\frac{1}{3}\Tr
\left\{\bW[\hat{\cS}_{i12}]-\bo\right\} \lbqa{3c} where
  $\pa\hat{\cS}_{i12}$ is the union of $\pa \cS_{i1}^{-1}$ and $\pa
  \cS_{i3}$. This is exactly the contribution of a quark traveling
  along line 3 and an anti-quark traveling along line 1=2. The spin
  contribution in the high energy limit of a quark and anti-quark is
  equal, namely $2s\de_{\la\la'}$, where $\la$ is the helicity in the
  initial and final state respectively. Thus in the limit $\al
  \rightarrow 0$ the baryon can be treated effectively as a meson, the
  point like di-quark traveling along line 1=2 replacing the
  anti-quark of the meson.

  After applying the non-Abelian Stokes theorem \ct{NonAb} the
  integration surface $S_{ij}$ of the Wegner-Wilson loop \beq
  W_{a'a}[\cS_{ij}]= \left[\cP_S\, e^{-ig\frac{1}{2}
    \int_{S_{ij}}\bF_{\mu\nu}(z,w)\,{\rm
      d}\si^{\mu\nu}(z)}\right]_{a'a} \lbq{WWloop} are the sliding
  sides of the pyramid belonging to quark $j$ of baryon $i$ ($\cP_S$
  stands for surface-ordered integration). For an illustration in the
  transversal plane see \fig{geometrie}.

  By averaging \eq{Jtilde} with wave functions with mean radius $S_i$,
  \beqa \hat{J}(\vec{b})&=&\int\,{\rm d^2}\vec{R}_1\int\,{\rm
    d^2}\vec{R}_2\,\tilde{J}(\vec{b},\vec{R}_1,\vec{R}_2)
  |\Ps(\vec{R}_1)|^2|\Ps(\vec{R}_2)|^2\nn\\ \Ps(\vec{R}_i)&=&
  \sqrt{\frac{2}{\pi}}\frac{1}{S_i}e^{-\frac{|\vec{R}_i|^2}{S_i^2}},
  \lbqa{T3B} the diffractive scattering amplitude $T$ at given center
  of mass energy $s$ is
\[
T= 2is \int \,{\rm d^2}\vec{b}\,\hat{J}(\vec{b},S_1,S_2).
\]
In order to calculate the scattering amplitude we first expand the
exponential of the Wegner-Wilson loops (\eq{WWloop}) in a power series
and plug this into \eq{Jtilde}. Then we express the parallel
transported field strengths in the adjoint basis of the $SU(3)$ and
perform the color-sums in \eq{Jtilde}. Finally we apply factorization
according the Gaussian model to the expectation value of the expanded
field strengths (for a detailed discussion see again \ct{9}) and
perform the surface integration using the MSV-correlator. It has been
shown that the surface integration over the correlator of a pair of
field strengths coming from the same baryon vanishes and the color-sum
is zero for only one field strength. So the leading contribution in
the expansion of the Wegner-Wilson loops comes from four field
strengths, two from baryon 1 and two from baryon 2. This contribution
has been calculated in \ct{9} and gives rise to a purely imaginary
$C$=+1 scattering amplitude.\\ Now we calculate the next contribution
where we have three field strengths from each baryon. There are three
possibilities (see \fig{3moeg}):\\ a) All three field strengths belong
to the same quark.\\ b) Two field strengths come from the same quark
and the third from another one.\\ c) To every quark belongs one field
strength.

The color-sums for these three possibilities are:\\ a) (e.g.~all three
fields strengths come from $q_{13}$) \beq \ep_{abc}\ep_{abc'}\left[
\bt^{C_1}\bt^{C_2}\bt^{C_3}\right]_{c'c}= 2\Tr\left[
\bt^{C_1}\bt^{C_2}\bt^{C_3}\right]= \frac{1}{2}d_{C_1 C_2
  C_3}+\frac{i}{2}f_{C_1 C_2 C_3} \lbq{colorsuma} b) (e.g.~two from
$q_{13}$ and one from $q_{12}$) \beq \ep_{abc}\ep_{ab'c'}\,\bt_{b'b}^B
\left[ \bt^{C_1}\bt^{C_2}\right]_{c'c}= -\frac{1}{4}d_{B C_1
    C_2}-\frac{i}{4}f_{B C_1 C_2} \lbq{colorsumb} \beq \mbox{c)
    }\ep_{abc}\ep_{a'b'c'}\,\bt_{a'a}^A\,\bt_{b'b}^B\,\bt_{c'c}^C=
\frac{1}{2}d_{ABC}
  \lbq{colorsumc} Here $f_{ABC}$ and $d_{ABC}$ are the structure
  constants and the symmetric $d$-symbols of $SU(3)$.\\ The next step
  is to factorize the expectation value into pairs. To avoid very long
  formulas we introduce the following shorthand notation:
\[
\int \int <i^Aj^B>\; :=\; \int_{S_{1i}}\int_{S_{2j}} <g^2 \,
F_{\mu\nu}^A(x,w)\,F_{\rh\si}^B(y,w)>\, {\rm d}\si^{\mu\nu}(x)\, {\rm
  d}\si^{\rh\si}(y)
\]
For example $\int \int <2^A3^B>$ stands for the surface integration
over the correlator with one field strength running over the pyramid
belonging to quark 2 of baryon 1 and the other running over the
pyramid belonging to quark 3 of baryon 2.\\ There are 7 permutations
for coupling the two baryons depending on which possibility of a), b)
or c) in \fig{3moeg} is chosen.\\ Let us start with possibility a) for
both baryons. Expanding e.g.~$W_{c'c}[\cS_{13}]$ and
$W_{f'f}[\cS_{23}]$ up to third order and using the color-sum
\eq{colorsuma} we get the following contribution to \eq{Jtilde}: \beqa
&&-\frac{1}{36}\left( \frac{-i}{2} \right)^6 \left[ \frac{1}{2}d_{C_1
  C_2 C_3}+\frac{i}{2}f_{C_1 C_2 C_3} \right] \left[ \frac{1}{2}d_{F_1
  F_2 F_3}+\frac{i}{2}f_{F_1 F_2 F_3} \right] \times \nn\\ && \cP_S
\int \cdots \int <3^{C_1}3^{C_2}3^{C_3}\;3^{F_1}3^{F_2}3^{F_3}> \enqa
Now we factorize the expectation value into pairs. The $d_{C_1 C_2
  C_3}\cdot f_{F_1 F_2 F_3}$ color structure vanishes because the
correlator (\eq{Korrelator}) is diagonal in color. We thus arrive at
the expression \beqa &\Rightarrow& -\frac{1}{36\cdot 4}\left(
\frac{-i}{2} \right)^6 \left[ d_{C_1 C_2 C_3}\cdot d_{F_1 F_2
  F_3}-f_{C_1 C_2 C_3}\cdot f_{F_1 F_2 F_3}\right] \times \nn\\ 
&&\cP_S \int \cdots \int \left\{
<3^{C_1}3^{F_1}><3^{C_2}3^{F_2}><3^{C_3}3^{F_3}>\, + \, \mbox{5
  permutations} \right\} \enqa The $d_{C_1 C_2 C_3}\cdot d_{F_1 F_2
  F_3}$ color-sum is the same for all six factorizations whereas the
$f_{C_1 C_2 C_3}\cdot f_{F_1 F_2 F_3}$ sum has different signs. In the
first structure we can therefore replace the surface-ordered integrals
by usual ones corrected with $1/3!$ for each baryon. Whereas in the
second structure we have to perform the surface-ordered integrals what
is technically very complicated. In this publication we concentrate on
the $C=-1$ contribution and we will show that for this contribution
only the $d\cdot d$ structure is needed.\\ If one baryon is replaced
by an anti-baryon we have to invert the orientation of the
corresponding Wegner-Wilson loops. This yields a factor $(-1)^3$.
Furthermore inverting the surface ordering of the three field
strengths gives an extra $(-1)$ for the $f$ structure whereas the $d$
structure is symmetric and remains unchanged. This shows that only the
$d\cdot d$ structure gives rise to a change in sign by replacing one
baryon by an anti-baryon.\\ We finally get, with $d_{C_1 C_2 C_3}\cdot
d_{C_1 C_2 C_3}=\frac{40}{3}$, \beqa &\Rightarrow& -\frac{1}{36\cdot
  4}\left( \frac{-i}{2} \right)^6 \frac{6}{3!\cdot 3!}
\frac{40}{3}\frac{1}{(N_C^2-1)^3}\left[ \int\int <3^A3^A>
\right]^3\nn\\ &=&-i\frac{5}{9\cdot12^3\cdot (N_C^2-1)^3 \cdot 36}
\;\tilde{\chi}_{33}^3 \enqa where
\[
\tilde{\ch}_{ij}:=\frac{-12i}{4}\int\int<i^Aj^A>.
\]
Taking into account all permutations for possibility a) we get: \beq
-i\frac{5}{9\cdot12^3\cdot (N_C^2-1)^3 \cdot 36} \;\left[
\tilde{\chi}_{11}^3+\tilde{\chi}_{12}^3+\tilde{\chi}_{13}^3+
\tilde{\chi}_{21}^3+\tilde{\chi}_{22}^3+\tilde{\chi}_{23}^3+
\tilde{\chi}_{31}^3+\tilde{\chi}_{32}^3+\tilde{\chi}_{33}^3\right]
\lbq{ergebnisaa} The calculation of all the remaining contributions
(all combinations of the three possibilities in \fig{3moeg}) is very
similar to the previous case so we only give the final result for the
$C$=$P$=--1 contribution to the reduced scattering amplitude
(\eq{Jtilde}): \beqa \tilde{J}^{C=-1}&=&-i\frac{5}{9\cdot12^3\cdot
  (N_C^2-1)^3 \cdot 36}\times\nn\\ &&\bigg[ \tilde{\chi}_{11}^3\, + \,
\mbox{8 per.} - \frac{3}{2} \tilde{\chi}_{11}^2\,\tilde{\chi}_{12}\, +
\, \mbox{35 per.}\nn\\ &&+\frac{3}{4}
\tilde{\chi}_{11}^2\,\tilde{\chi}_{22}\, + \, \mbox{35 per.} \, +
\,\frac{3}{2}
\tilde{\chi}_{11}\,\tilde{\chi}_{12}\,\tilde{\chi}_{22}\, + \,
\mbox{35 per.}+6
\tilde{\chi}_{11}\,\tilde{\chi}_{12}\,\tilde{\chi}_{13}\, + \, \mbox{5
  per.}\nn\\ &&-3
\tilde{\chi}_{11}\,\tilde{\chi}_{12}\,\tilde{\chi}_{23}\, + \,
\mbox{35 per.} +6
\tilde{\chi}_{11}\,\tilde{\chi}_{22}\,\tilde{\chi}_{33}\, + \, \mbox{5
  per.} \bigg].  \lbqa{ergebniss} The evaluation of the
$\tilde{\ch}_{ij}$ follows the same lines as in reference \ct{9}.

{\Large \bf 3 Numerical results}\hspace{1cm}Using \eq{ergebniss},
\eq{T3B} and the results for ${\rm Im}\left[ T^{p p}(s,0)\right]$ we
compute the leading contribution to the rho parameter.
\[
\De \rh(s) =\frac{{\rm Re}\left[ T^{\bar p p}(s,0)\right]}{{\rm
    Im}\left[ T^{\bar p p}(s,0)\right]}-\frac{{\rm Re}\left[ T^{ p
    p}(s,0)\right]}{{\rm Im}\left[ T^{ p p}(s,0)\right]}=-2\frac{{\rm
    Re}\left[ T^{ p p}(s,0)\right]}{{\rm Im}\left[ T^{ p
    p}(s,0)\right]}
\]
The results depend on the geometry chosen for the baryon and on the
parameters of the MSV, that is the correlation length $a$ and the
condensate $<g^2FF>$. In the same way as it has been done in \ct{9}
the size of the proton, $a$ and $<g^2FF>$ have been determined in
\ct{13} for the star-like and a linear geometry (see \fig{ort}). Here
a somewhat more complete expression for the MSV-correlator
(\eq{Korrelator}) has been used leading to a minor change of
parameters as compared to \ct{9}. At $\sqrt{s}=541$ GeV we found:

\begin{table}[ht]
  \centerline{
\begin{tabular}{|c||c|c|}\hline
  &star-like&linear\\ \hline\hline $<g^2FF>$&1.88 $\GeV ^4$&3.07 $\GeV
  ^4$\\ \hline $a$&0.371 fm&0.332 fm\\ \hline $S_{\rm proton}$&1.93
  $a$&2.62 $a$\\ \hline
\end{tabular}}
\caption{The two parameter sets}
\label{parameter}
\end{table}

With this set of parameters we present in \fig{plots} $\De \rh$ at
$\sqrt{s}=541$ GeV as a function of $r_\perp$ and $r_\|$ (for
illustration see \fig{ort}).

{\Large \bf 4 Discussion}\hspace{1cm}As can be seen from \fig{plots} a
clustering of two quarks to a di-quark with a radius smaller or equal
to $0.3$ fm yields already a drastic suppression of $\De \rh$ to a
value $|\De \rh | \le 0.02$ which is compatible with the analysis of
experiments.\\ It should be noted that even for a meson or baryon in
the di-quark picture the contribution of the $C$=$P$=--1 exchange is
appreciable for a given Wegner-Wilson loop. But constructing the
hadrons by averaging the loops with wave functions cancels these
contributions. For a clustering to a di-quark with finite radius the
$C$=$P$=--1 contribution is suppressed but not completely canceled.\\ 
There is plenty of other evidence for di-quark clustering in baryons:
The scaling violation in nucleon structure functions \ct{Ans}, the
strong attraction in the scalar di-quark channel in the instanton
vacuum \ct{Shur} and the $\De I=\frac{1}{2}$ enhancement in semi
leptonic decays of baryons \ct{Dos}. Scaling violation in deep
inelastic scattering is sensible to the form factors of the di-quark,
which is modeled by a pole fit with a pole mass of $\sqrt{3}$ to
$\sqrt{10}$ GeV. This corresponds to di-quark radii of 0.3 to 0.16 fm.
They are according to our model sufficiently small to give a
suppression of $\De \rh$ to values below 0.02 even for the transversal
extension. If the nucleon has a linear structure no suppression by
di-quark clustering is necessary at all.\\ Our calculation have been
performed in a specific non-perturbative model. But since the limiting
case of a vanishing di-quark radius leads quite generally to an
odderon cancelation as in the case of mesons (see \eq{3c} and the
discussion of it) we think that the suppression of the $C$=$P$=--1
exchange is generally to be caused by the structure of the nucleon and
cannot be seen on the quark level.

{\Large \bf Acknowledgments}\hspace{1cm}The authors thank G.~Kulzinger, 
P.~V.~Landshoff and O.~Nachtmann for discussions and valuable comments.

\epsfxsize8.5cm
\befi{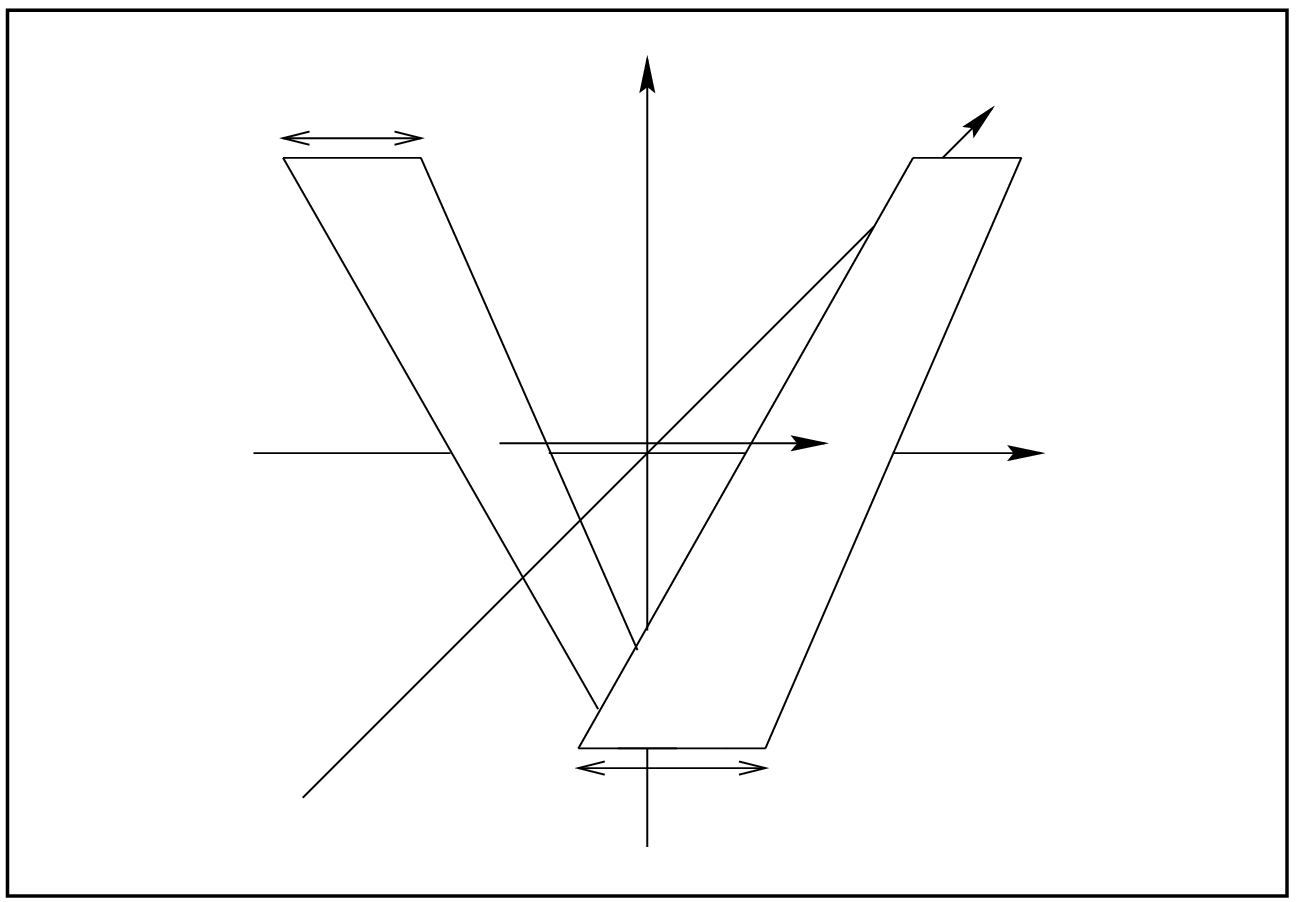}
\unitlength.85cm
\begin{picture}(0,0)
\put(-8.5,4.5){loop 1}
\put(-2,4.5){loop 2}
\put(-2.2,3){$\vec{x}$}
\put(-4.9,6.2){$x^0$}
\put(-2.2,6){$x^3$}
\put(-7.5,6.1){$\vec{R}_1$}
\put(-4.9,.6){$\vec{R}_2$}
\put(-5.4,3.7){$\vec{b}$}
\end{picture}
\caption{Two loops with transversal extension $\vec{R}_1$ and 
$\vec{R}_2$ and light-like sides. Loop 1 describes a colorless 
$q\bar{q}$-pair running 
in negative 3-direction and loop 2 in positive 3-direction. The 
impact parameter $\vec{b}$ is chosen to be purely transverse.}
\lbfi{2loops} \epsfxsize8.5cm
\befi{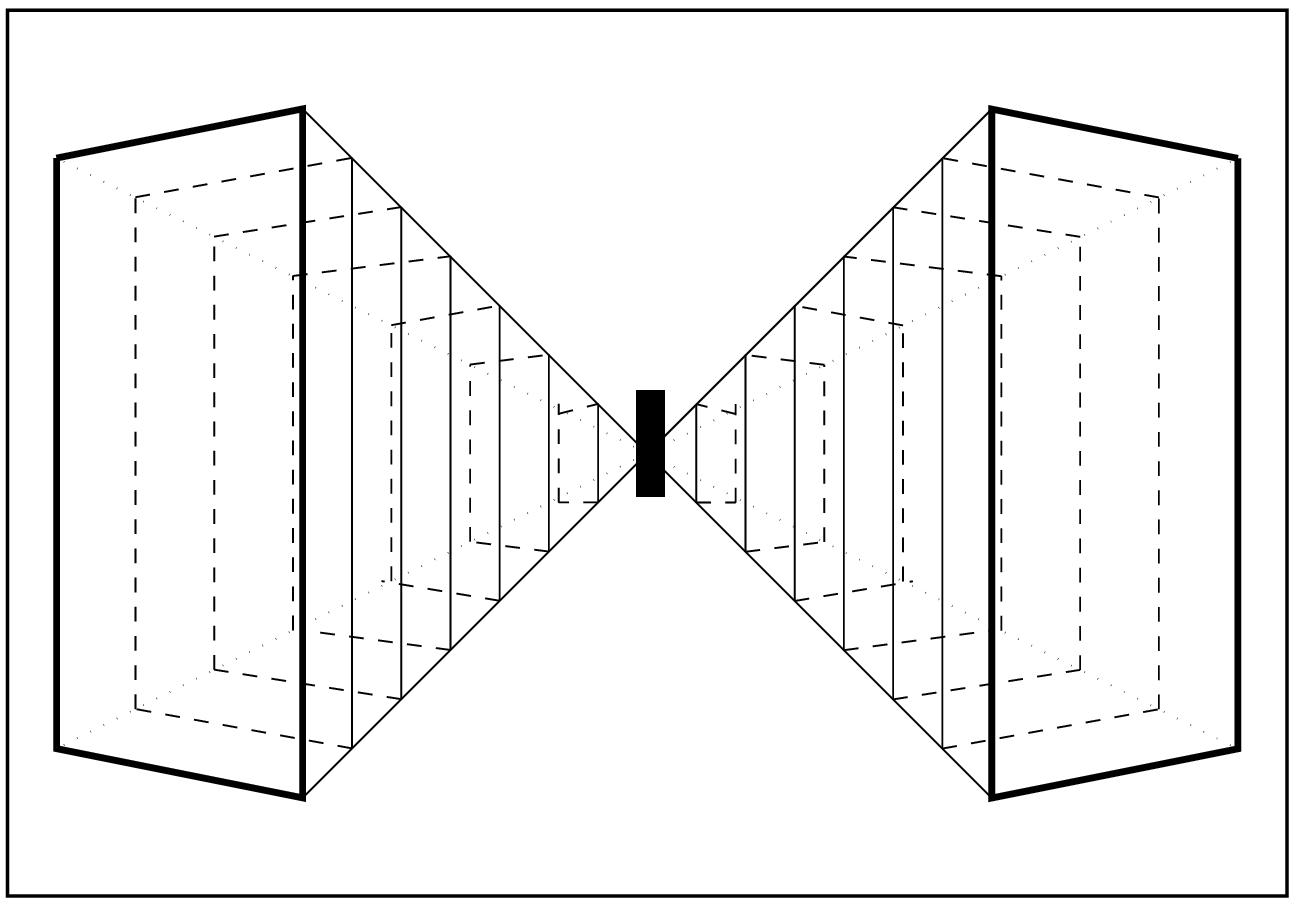}
\unitlength.85cm
\begin{picture}(0,0)
\put(-9,6.3){loop 1}
\put(-2,6.3){loop 2}
\put(-5.1,3.7){$C$}
\end{picture}
\caption{Somewhat tilted view of the loops after applying the 
non-Abelian Stoke's theorem with the reference point $C$ which 
is common to both surfaces.}
\lbfi{2pyramiden} \epsfxsize6cm
\befi{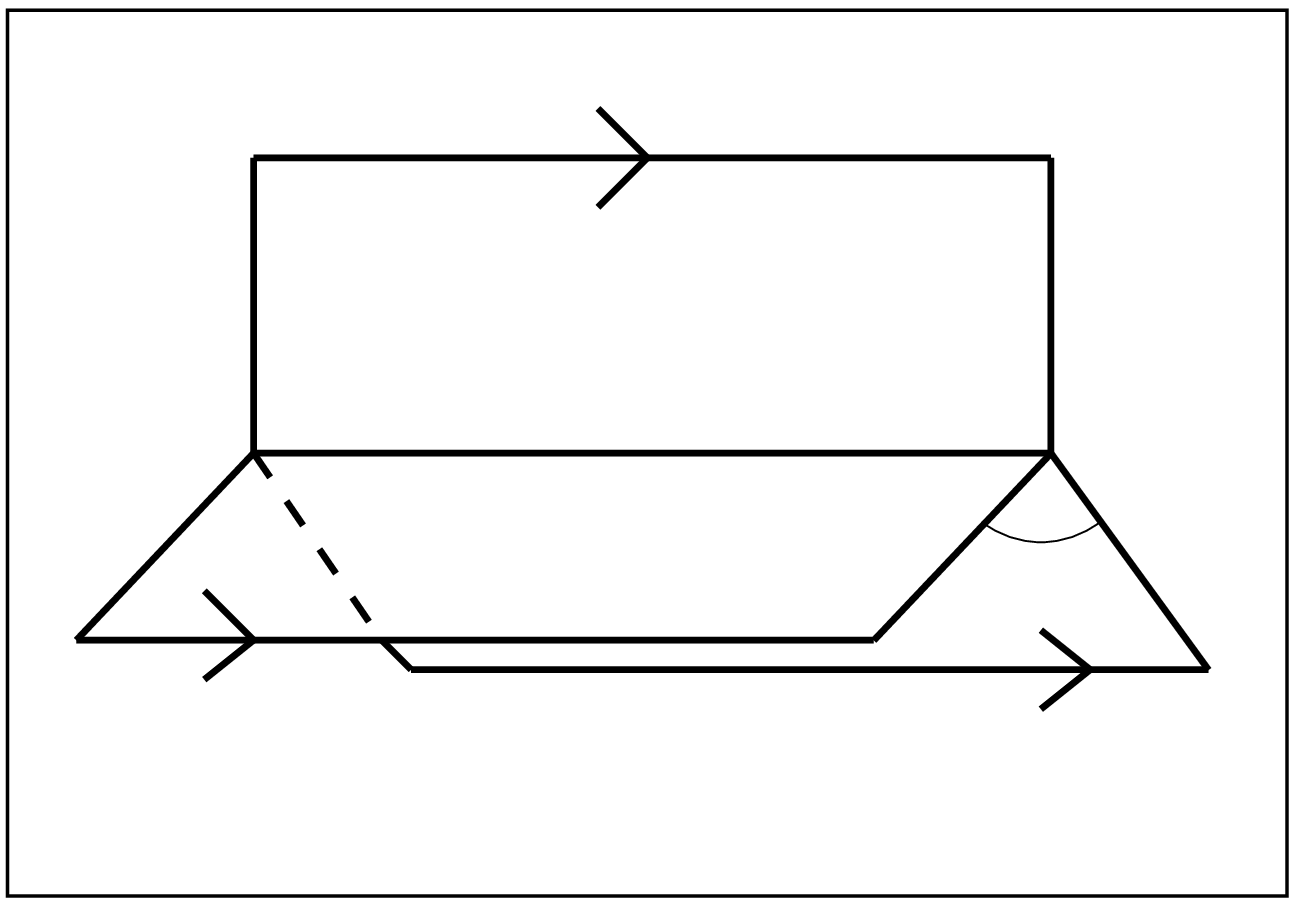}
\unitlength.813cm
\begin{picture}(0,0)
\put(-1.6,2.2){$\al$}
\put(-1.6,.7){$\pa \cS_{i2}$}
\put(-6.6,.9){$\pa \cS_{i1}$}
\put(-3.6,4.5){$\pa \cS_{i3}$}
\end{picture}
\caption{A baryon is constructed out of 3 loops with one common 
line which transforms like a color singlet. Here $\pa \cS_{ij}$ 
denotes the loop corresponding to quark j of baryon i.}
\lbfi{3loops} \epsfxsize8.5cm
\befi{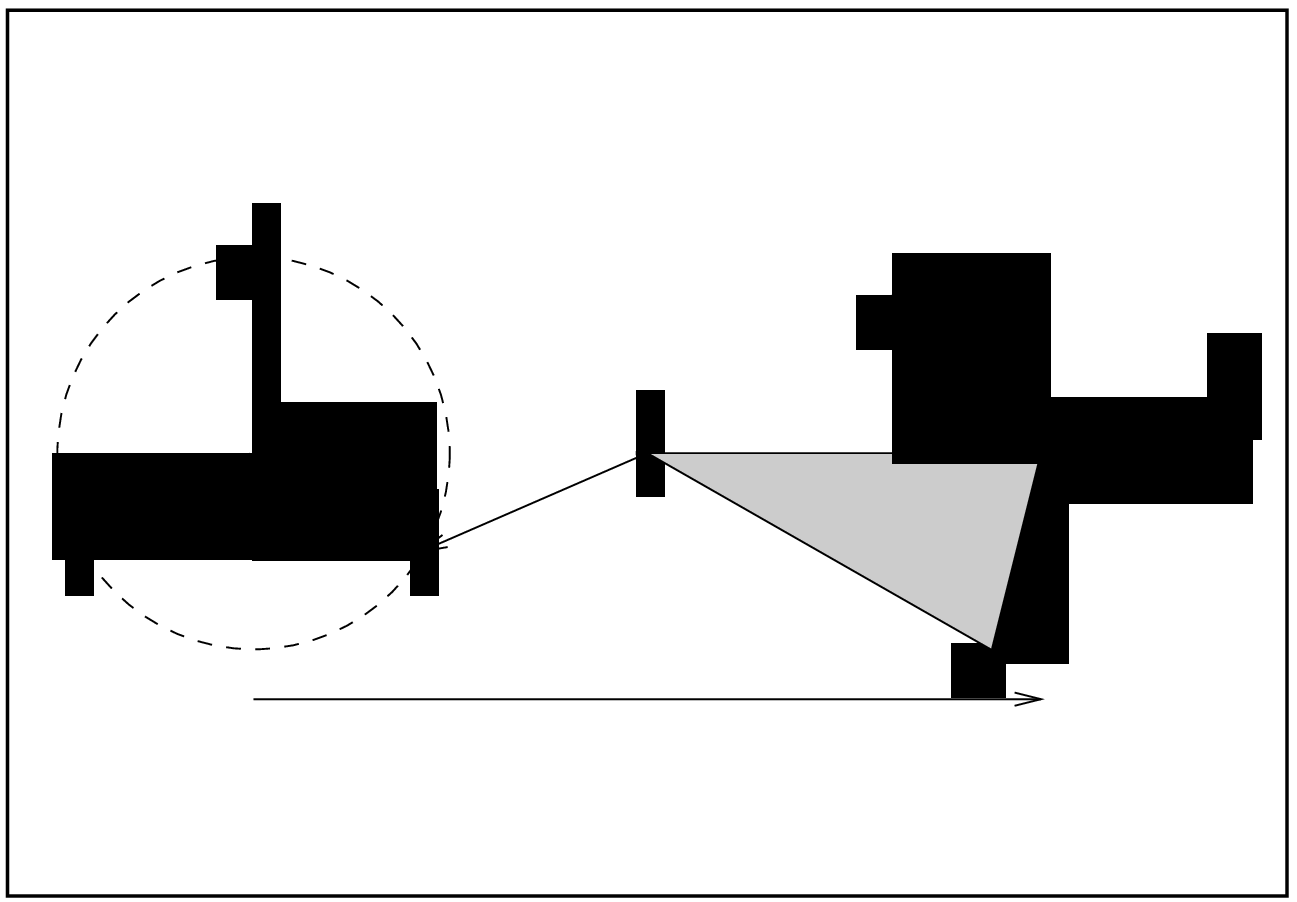}
\unitlength.654cm
\begin{picture}(0,0)
\put(-11.5,7.5){baryon 1}
\put(-3.5,7.5){baryon 2}
\put(-10.7,7){$q_{11}$}
\put(-11.3,5.5){$\vec{R}_1$}
\put(-3.1,5.5){$\vec{R}_2$}
\put(-12.7,3){$q_{12}$}
\put(-8.7,3){$q_{13}$}
\put(-4.2,6.5){$q_{21}$}
\put(-2.7,2.5){$q_{22}$}
\put(-.9,4.5){$q_{23}$}
\put(-8.2,4.3){$\vec{r}_{13}$}
\put(-7,4.8){$(0,0)$}
\put(-6.9,2.3){$\vec{b}$}
\put(-4.5,3.8){$S_{22}$}
\end{picture}
\caption{The geometry of the two baryons in the transversal plane. 
The vector pointing to quark $j$ of baryon $i$ ($q_{ij}$) is 
denoted by $\vec{r}_{ij}$ and the vector $\vec{R}_i$ points to 
quark $q_{i1}$. $S_{22}$ is the projection on the transversal 
plane of the integration surface for $W_{ee'}[\cS_{22}]$. In this 
picture the three quarks of a baryon are arranged like a star with 
equal angles, but the calculation is valid for any geometry.}
\lbfi{geometrie} \epsfxsize8cm
\befi{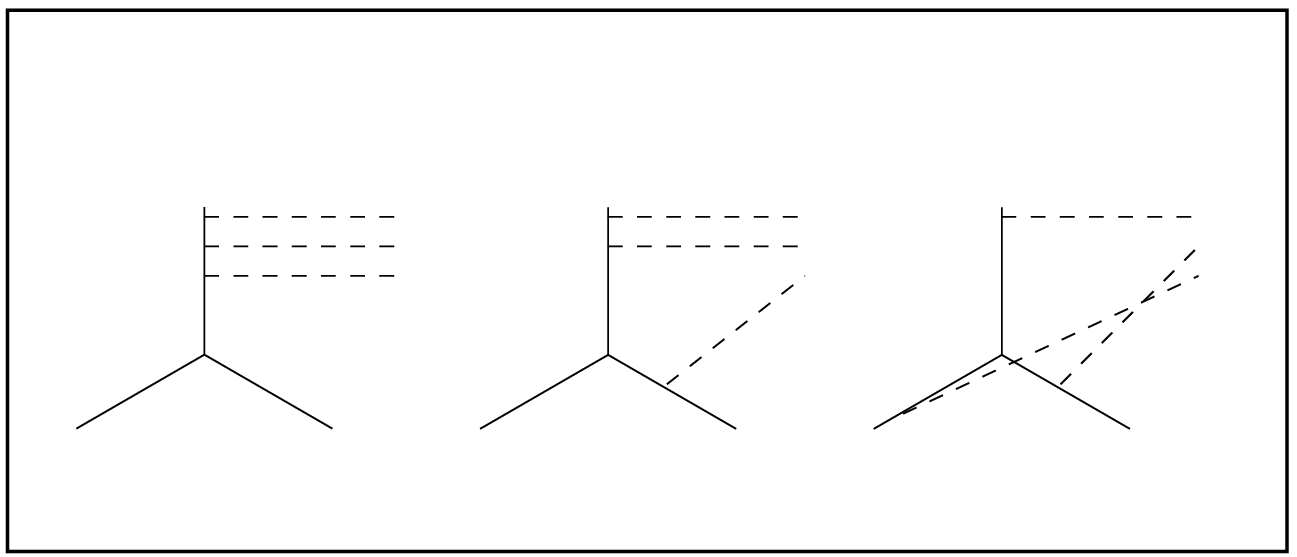}
\unitlength.615cm
\begin{picture}(0,0)
\put(-12,4){possibility a)}
\put(-8,4){possibility b)}
\put(-4,4){possibility c)}
\end{picture}
\caption{The three possibilities for arranging the three field 
strengths on the three quarks.}
\lbfi{3moeg} \epsfxsize8.5cm
\befi{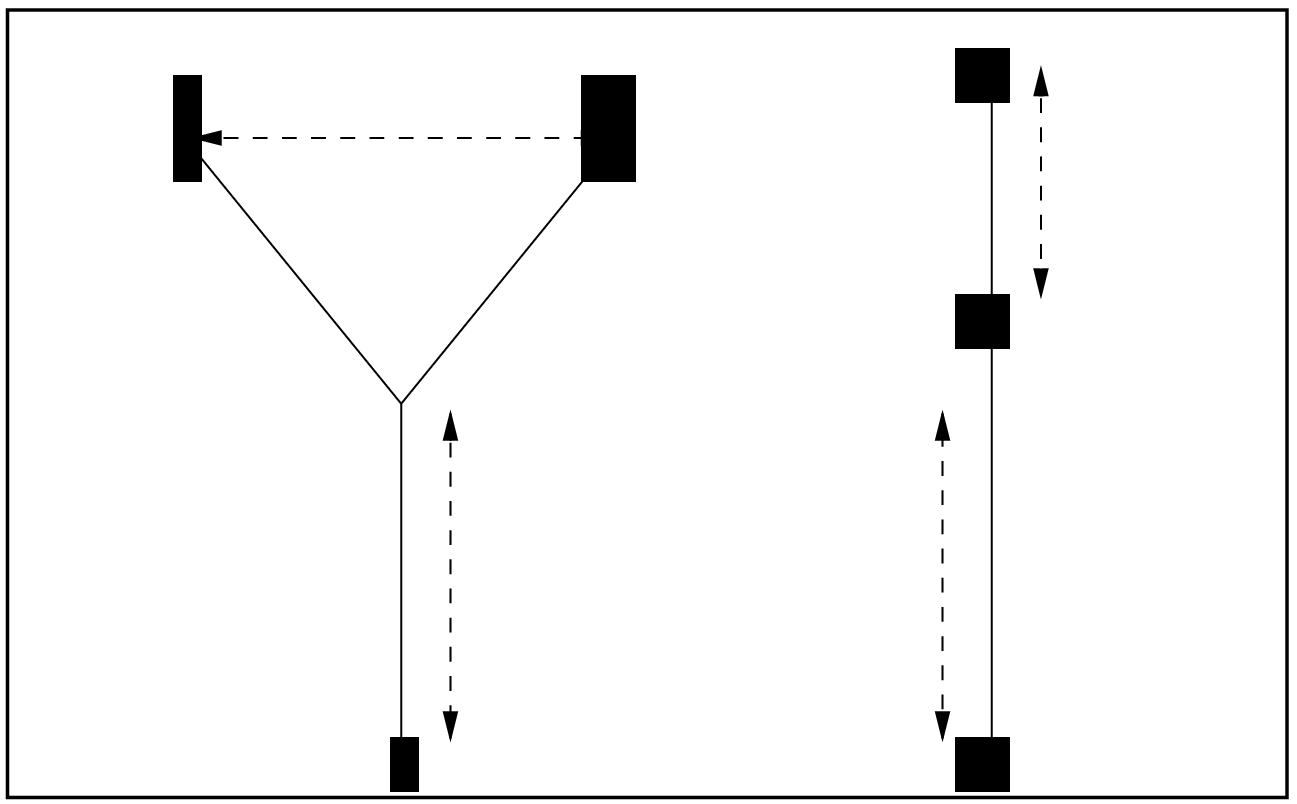}
\unitlength.654cm
\begin{picture}(0,0)
\put(-8.2,2.2){$S$}
\put(-4,2.2){$S$}
\put(-9.2,7){$r_\perp$}
\put(-2.4,6.2){$r_\|$}
\end{picture}
\caption{In this figure we show the two different geometries chosen 
for the baryon. The proton radius is given in table 1
and for small quark distances $r_\perp$ and $r_\|$ we have a 
di-quark configuration.}
\lbfi{ort}
\begin{figure}[ht]
\begin{minipage}{7.5cm}
  \epsfxsize7.5cm
\epsffile{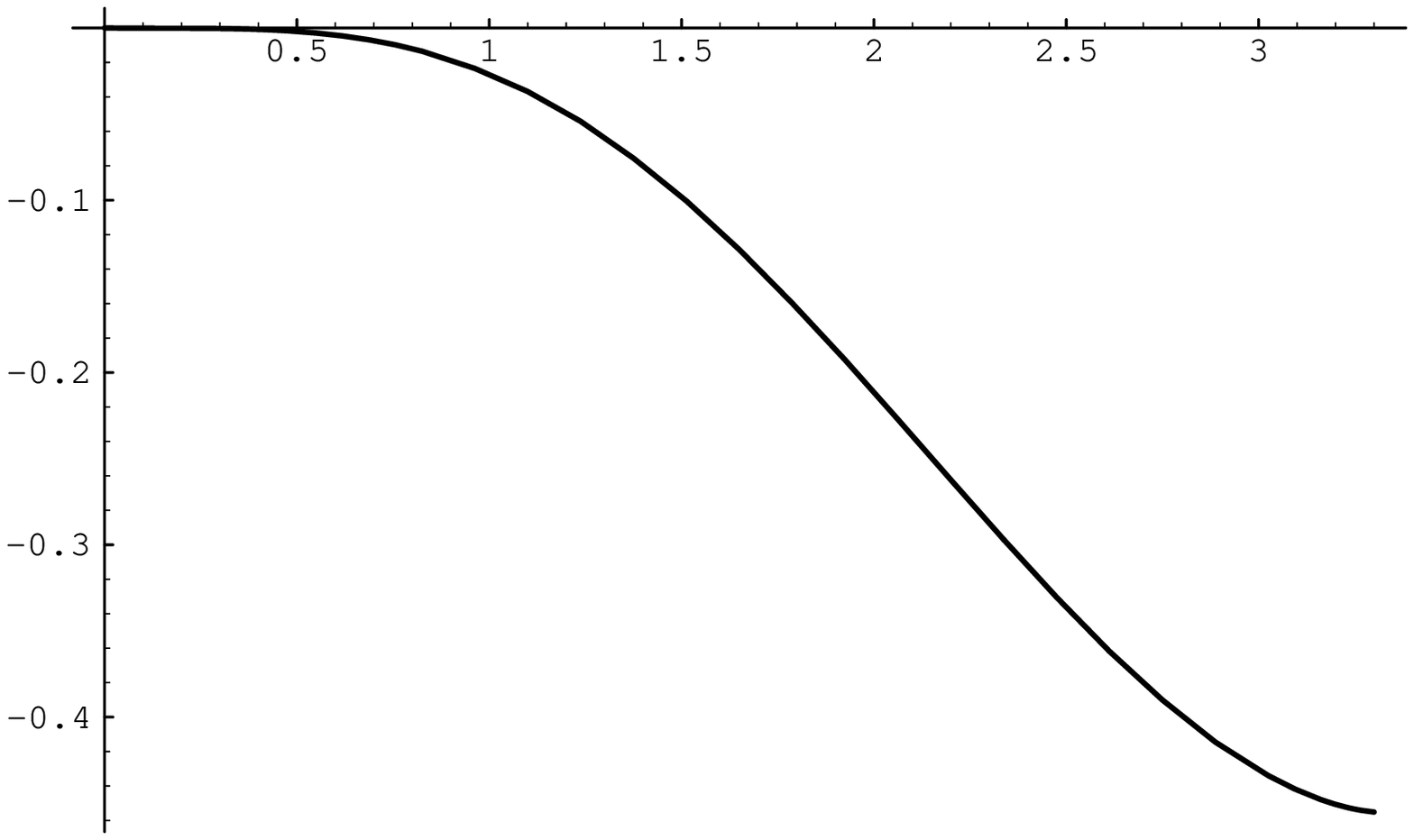}
\unitlength1cm
\begin{picture}(0,0)
\put(7,4.3){$\frac{r_\perp}{a}$}
\put(.7,.7){$\De \rh$}
\end{picture}
\begin{center}
  star-like geometry\\ ( $a$=0.371 fm, $S$=1.93$a$ )
\end{center}
\end{minipage}\hfill
\begin{minipage}{7.5cm}
  \epsfxsize7.5cm
\epsffile{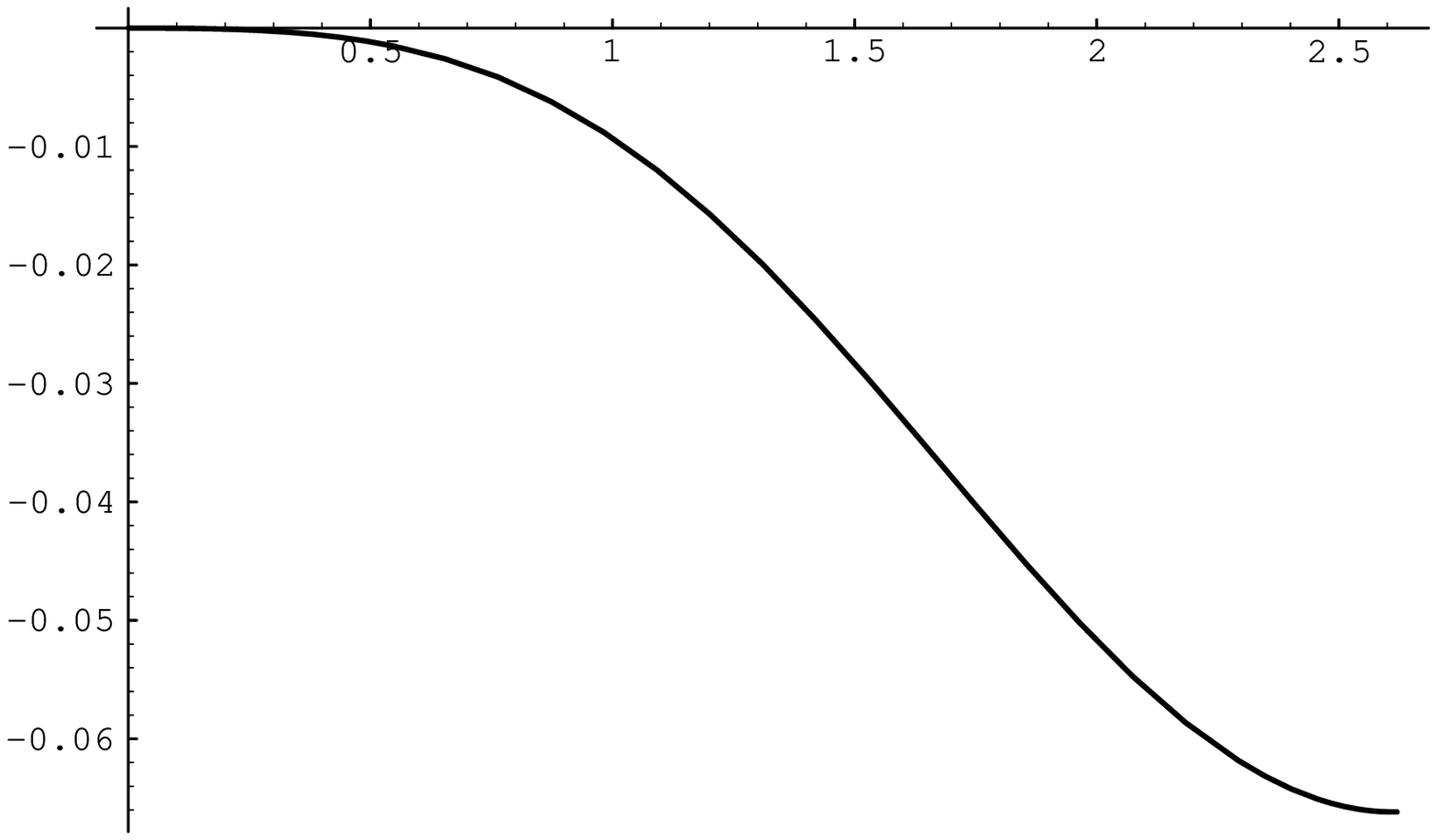}
\unitlength1cm
\begin{picture}(0,0)
\put(7,4.3){$\frac{r_\|}{a}$}
\put(.7,.7){$\De \rh$}
\end{picture}
\begin{center}
  linear geometry\\ ( $a$=0.332 fm, $S$=2.62$a$ )
\end{center}
\end{minipage}
\caption{$\De \rh$ at UA4/2 energies for proton-(anti)proton 
scattering as a function of the quark distances $r_\perp$ and 
$r_\|$. Note the different scales.}
\label{plots}
\end{figure}

\begin{thebibliography}{12}
\bibitem{1} L.~Lukaszuk, B.~Nicolescu, {\it Lett.Nuov.Cim.}{\bf 8}
  (1973) 405 \bibitem{2} UA4 Collaboration (D.~Bernard et.al.~), {\it
    Phys.Lett.}{\bf B198} (1987) 583 \bibitem{3} UA4/2 Collaboration
  (D.~Bernard et.al.~), {\it Phys.Lett.}{\bf B316} (1993) 448
\bibitem{4} C.~Bourelly, A.~Martin, Proc.~of the LHC workshop,
  Lausanne (1984)\\ A.~Donnachie, P.~Landshoff, {\it Nucl.Phys.}{\bf
    B244} (1984) 322\\ C.~Bourelly, J.~Soffer, T.~T.~Wu, {\it
    Phys.Lett.}{\bf B196} (1987) 237\\ P.~Gauron, E.~Leader,
  B.~Nicolescu, {\it Nucl.Phys.}{\bf B299} (1988) 640\\ P.~Kroll,
  W.~Schweiger, {\it Nucl.Phys.}{\bf A503} (1989) 865\\ 
  R.~J.~M.~Covolan et al., {\it Z.Phys.}{\bf C58} (1993) 109
\bibitem{LaPo} A.~Donnachie, P.~V.~Landshoff, {\it Phys.Lett.}{\bf
    B296} (1992) 227 \bibitem{5} P.~Gauron, Proc.~V.th Blois workshop,
  ed.~Fried et al.~Singapore 1994 \bibitem{6} A.~P.~Contogouris, V.th
  Blois workshop, ed.~Fried et al.~Singapore 1994 \bibitem{7}
  P.~V.~Landshoff, O.~Nachtmann, {\it Z.Phys.}{\bf C35} (1987) 405
\bibitem{8} A.~Donnachie, P.~V.~Landshoff, {\it Nucl.Phys.}{\bf B348}
  (1991) 297 \bibitem{9} H.~G.~Dosch, E.~Ferreira, A.~Kr\"amer, {\it
    Phys.Rev.}{\bf D50} (1994) 1992 \bibitem{10} O.~Nachtmann, {\it
    Ann.Phys.}{\bf 209} (1991) 436 \bibitem{11} H~.G.~Dosch, {\it
    Phys.Lett.}{\bf B190} (1987) 177 \bibitem{12} H.~G.~Dosch,
  Y.~A.~Simonov, {\it Phys.Lett.}{\bf B205} (1988) 339 \bibitem{NonAb}
  N.~E.~Bralic, {\it Phys.Rev.}{\bf D22} (1980) 3090\\ Y.~A.~Simonov,
  {\it Sov.J.Nucl.Phys.}{\bf 50} (1989) 134 \bibitem{13} H.~G.~Dosch,
  E.~Ferreira, G.~Kulzinger, M.~Rueter, in preparation \bibitem{Ans}
  M.~Anselmino, E.~Predazzi, S.~Ekelin, D.~B.~Lichtenberg, {\it
    Rev.Mod.Phys.}{\bf 65} (1993) 1199 \bibitem{Shur} T.~Sch\"afer,
  E.~V.~Shuryak, J.~J.~M.~Verbaarschot, {\it Nucl.Phys.}{\bf B142}
  (1994) 143 \bibitem{Dos} H.~G.~Dosch, M.~Jamin, B.~Stech, {\it
    Z.Phys.}{\bf C42} (1989) 167
\end{thebibliography}
\end{document}